\documentclass[11pt]{article}

\usepackage{amsmath,amssymb}

\usepackage{amsfonts}

\usepackage{amssymb,color}

\makeatletter \@addtoreset{equation}{section} \makeatother

\begin{document}

\title{On solutions of the
Pauli equation in non-static de Sitter metrics}

\author{E. M. Ovsiyuk, K. V. Kazmerchuk
\\Mozyr State Pedagogical University named after I.P. Shamyakin
\\e.ovsiyuk@mail.ru, kristinash2@mail.ru}

\maketitle

\begin{abstract}

A particle with spin 1/2 is investigated both in  expanding and
oscillating  cosmological de Sitter models. It is shown that these
space-time geometries admit  existence of the non-relativistic
limit in the  covariant Dirac equation. Procedure for transition
to the Pauli approximation is conducted in the equations in the
variables $(t, r)$, obtained after separating the angular
dependence of $(\theta, \phi)$ from  the wave function. The
non-relativistic systems of equations in the variables $(t, r)$ is
solved exactly in both models. The constructed wave functions do
not represent stationary states with fixed energy, however the
corresponding  probability density does not depend on the time.

\end{abstract}

\maketitle

\section{Introduction}

De Sitter and anti de Sitter models take attraction during long
time in the context of development of the quantum theory in curved
space-time. In particular, long history has the task of exact solving the wave equations
 for fields with different spins
 \cite{{Dirac-1935}}--\cite{RUDN-2012}; the main attention  was focused of relativistic wave equations.

In the present paper we will examine  the nonrelativistic Pauli
approximation for spin 1/2 particle in non-static metrics of de
Sitter models: we demonstrate how transition to Pauli
approximation can be performed in non-static de Sitter's metrics,
and we will construct exact solutions for these non-relativistic
systems in  both de Sitter models: expanding and oscillating ones.

\section{Pauli equation in expanding de Sitter Universe
}

Generally covariant Dirac equation in orthogonal coordinates
\cite{Red'kov-2009}
\begin{eqnarray}
 [  i \gamma^{k} \;  (\; e_{(k)}^{\alpha} \partial_{\alpha}
\; + \; B_{k}  \; )\; \;
 -  \; m  ] \;  \Psi  = 0               \; ,
\nonumber
\end{eqnarray}
\begin{eqnarray}
B_{k}(x) =
  {1 \over 2}  \; e_{(k)
   ;\alpha }^{\alpha}(x) =
  {1 \over 2} {1 \over \sqrt{-g}}  {\partial \over \partial x^{\alpha}}\sqrt{-g}e_{(k)}^{\alpha} \;
\nonumber
\end{eqnarray}

\noindent being specified for the non-static de Sitter metric
\begin{eqnarray}
x^{\alpha } = ( t , r , \theta  , \phi  ), \quad \quad
  dS^{2} =  dt^{2}
  -  \cosh^{2} t  [
dr^{2} + \sin^{2} r ( d\theta ^{2} + \sin^{2} \theta  d\phi^{2}
)  ]
\nonumber
\end{eqnarray}

\noindent and the corresponding tetrad
\begin{eqnarray}
e^{\alpha}_{(0)} = ( 1 , 0 , 0 , 0 ) \; ,
\quad
 e^{\alpha }_{(3)}
= \left ( 0 , { 1 \over   \cosh t } , 0 , 0 \right  ) \;
  ,
\nonumber
\\
e^{\alpha }_{(2)} =  \left ( 0 ,0 , 0 , {1 \over \cosh t  \sin r
\; \sin \theta} \right )\; , \quad
 e^{\alpha }_{(1)} = \left ( 0 , 0 ,{ 1 \over   \cosh
t \sin r}, 0  \right )\;
\nonumber
\end{eqnarray}

\noindent takes the form
\begin{eqnarray}
\left  [ i \gamma^{0}  \cosh t  \left (\;  \partial_{t} \; + \; {3
\over 2} \tanh t  \; \right )+
 {1 \over \sin r} \Sigma_{\theta \phi}
+ i \gamma^{3}  \left (   \partial_{r} \; + \;  {1
\over \tan r}  \;\right  )  -  \; m  \cosh t  \right ] \Psi  = 0
\,,
\nonumber
\end{eqnarray}

\noindent where
\begin{eqnarray}
\Sigma _{\theta ,\phi } \; = \;  i\; \gamma ^{1} \partial
_{\theta} \;+\; \gamma ^{2} \;  {\;  i \partial _{\phi} \; + \;
 i\; \sigma ^{12} \cos \theta   \over \sin \theta } \;  .
\nonumber
\end{eqnarray}

\noindent Separating a simple factor  in
the wave function
\begin{eqnarray}
\Psi (x)  = {1 \over \sin r} {1 \over \cosh ^{3/2} t } \; \varphi
(x) \;,
\nonumber
\end{eqnarray}

\noindent we  obtain a simpler equation
\begin{eqnarray}
\left [ i \gamma^{0}  \cosh t  \;  {\partial \over \partial t} + i
\gamma^{3} { \partial \over \partial r} +
 {1 \over \sin r} \Sigma_{\theta \phi}  -  \; m  \cosh t  \right ]   \varphi  = 0               \; .
\nonumber
\end{eqnarray}

 To diagonalize the total angular moment on the wave functions, we are to use the following substitution
 \cite{Book-2}
 (Wigner's functions \cite{1975-Varshalovich-Moskalev-Hersonskiy} are referred as
  $D^{j}_{-m,\sigma }(\phi ,\theta ,0)  \equiv
D_{\sigma}$):
\begin{eqnarray}
\varphi_{jm}(x)   = \left | \begin{array}{l}
        f_{1}(t,r) \; D_{-1/2} \\ f_{2}(t,r) \; D_{+1/2}  \\
        f_{3}(t,r) \; D_{-1/2} \\ f_{4}(t,r) \; D_{+1/2}
\end{array} \right | \; .
\label{1.4a}
\end{eqnarray}

\noindent With the use of recurrent relations
\cite{1975-Varshalovich-Moskalev-Hersonskiy}
\begin{eqnarray}
\partial_{\theta} \; D_{+1/2} \; = \;  a\; D_{-1/2}  - b \; D_{+3/2}  \; ,
\nonumber
\\
{- m - 1/2   \cos \theta  \over  \sin \theta }  D_{+1/2} =
- a   D_{-1/2} - b   D_{+3/2}   ,
\nonumber
\\
\partial_{\theta} \;  D_{-1/2} \; = \;  b \; D_{-3/2} - a \; D_{+1/2}  \; ,
\nonumber
\\
{- m + 1/2  \cos \theta \over \sin \theta}    D_{-1/2}  =
 - b  D_{-3/2}  - a   D_{+1/2}   ,
\nonumber
\label{1.4b}
\end{eqnarray}

\noindent where
\begin{eqnarray}
a = {j +  1/2 \over 2} , \quad b  = {1\over 2}
\,\sqrt{(j-1/2)(j+3/2)}\,,
\nonumber
\end{eqnarray}

\noindent we find the  action of the angular operator
 $\Sigma_{\theta
\phi}$
\begin{eqnarray}
 \Sigma _{\theta ,\phi } \, \varphi  _{ jm}(x) \,  = \, i\, \nu \,
 \left | \begin{array}{r}
        - \, f_{4}(t,r) \, D_{-1/2}  \\  + \, f_{3}(t,r) \, D_{+1/2} \\
        + \, f_{2}(t,r) \, D_{-1/2}  \\  - \, f_{1}(t,r) \, D_{+1/2}
\end{array} \right | ,
\label{1.4c}
\end{eqnarray}

\noindent  where $ \nu  = j + 1/2.$ Then produce four equations in
variables $(t,r)$:
\begin{eqnarray}
i \cosh t { \partial \over \partial t}    f_{3}
  -  i  {\partial \over \partial r}  f_{3}  - i {\nu \over \sin r}
f_{4}
     -  m \cosh t  f_{1} =   0   ,
\nonumber
\\
i \cosh t { \partial \over \partial t}   f_{4} +  i  {\partial
\over \partial r} f_{4}  + i {\nu \over \sin r} f_{3}
   -  m
\cosh t  f_{2} =   0     ,
\nonumber
\\
i \cosh t { \partial \over \partial t}   f_{1} +  i  {\partial
\over \partial r} f_{1}  + i {\nu \over \sin r} f_{2}
  -  m
\cosh t f_{3} =   0   ,
\nonumber
\\
i \cosh t { \partial \over \partial t}    f_{2}
  -  i  {\partial \over \partial r} f_{2}- i {\nu \over \sin r} f_{1}
    -  m \cosh t f_{4} =   0   .
\label{1.5}
\end{eqnarray}

\noindent Usual $P$-reflection operator in Cartesian basis of the tetrad
$\hat{\Pi}_{C}  =  i \gamma ^{0} \otimes \hat{P}$
\begin{eqnarray}
\hat{\Pi}_{C.}  = \left | \begin{array}{cccc}
          0 &  0 &  i &   0  \\
          0 &  0 &  0 &   i  \\
          i &  0 &  0 &   0  \\
          0 &  i &  0 &   0
\end{array} \right |       \; \otimes  \; \hat{P} \; ,\quad
\hat{P} (\theta , \phi ) = (\pi  - \theta, \; \phi+ \pi )
\nonumber
\end{eqnarray}

\noindent after translating it to spherical tetrad
 will take the form
\begin{eqnarray}
\hat{\Pi}_{sph} \;  = \left | \begin{array}{cccc}
0 &  0 &  0 & -1   \\
0 &  0 & -1 &  0   \\
0 &  -1&  0 &  0   \\
-1&  0 &  0 &  0
\end{array} \right |
\; \otimes  \; \hat{P} \; .
\label{1.6}
\end{eqnarray}

\noindent From eigenvalue equation $ \hat{\Pi}_{sph}
\Psi _{jm} =  \Pi \; \Psi _{jm}$ with the us of the known identity
\cite{1975-Varshalovich-Moskalev-Hersonskiy}
\begin{eqnarray}
\hat{P}   \; D^{j}_{-m,\sigma } (\phi ,\theta ,0)  = (-1)^{j} \;
D^{j}_{-m,-\sigma} (\phi ,\theta ,0)
\nonumber
\end{eqnarray}

\noindent   we derive linear restrictions of the functions:
\begin{eqnarray}
\Pi = \, \delta \,  (-1)^{j+1} , \;\; \delta  = \pm 1 \, ,
\quad
f_{4} = \, \delta \,  f_{1} , \quad   f_{3} = \,\delta \, f_{2} \,
.
\nonumber
\end{eqnarray}

\noindent Thus, the wave function for states with fixed parity
takes
 the form
\begin{eqnarray}
\varphi (x)_{jm\delta } \; =
 \left | \begin{array}{r}
     f_{1}(t,r) \; D_{-1/2} \\
     f_{2}(t,r) \; D_{+1/2} \\
\delta \; f_{2}(t,r) \; D_{-1/2}   \\
\delta \; f_{1}(t,r) \; D_{+1/2}
\end{array} \right |   .
\label{1.8}
\end{eqnarray}

Allowing for  (\ref{1.8}), from  (\ref{1.5}) we get more simple equations
\begin{eqnarray}
\left ( {\partial \over \partial r} + {\nu \over \sin r} \right )
f
+ \left  ( i  \cosh t\, { \partial \over \partial t} +
 \delta  m  \cosh t \right ) g  = 0  ,
 \nonumber
\\
\left ({\partial  \over \partial r}
 - {\nu \over \sin r} \right )
g
 - \left  ( i  \cosh t\, {\partial  \over \partial t}  -
\delta  m  \cosh t \right ) f = 0      ,
\label{1.9}
\end{eqnarray}

\noindent where instead of   $f_{1}(t,r) $  and   $f_{2}(t,r) $  their linear combinations are used:
\begin{eqnarray}
f (t,r)  =  {f_{1} + f_{2} \over \sqrt{2}} \; , \qquad g (t,r)  =
{f_{1} - f_{2} \over i \sqrt{2}} \; .
\nonumber
\end{eqnarray}

Now, let us perform nonrelativistic approximation in the system
 (\ref{1.9}) (here we will adhere the method exposed in \cite{Red'kov-2009, Book-2}).
First, we separate the rest energy
by the formal change
\begin{eqnarray}
i {\partial \over \partial t} \quad  \Longrightarrow \quad  M + i
{\partial \over \partial t}   \,.
\nonumber
\end{eqnarray}

The cases  $\delta = \pm 1$ should be considered separately:

$\delta = +1$,
\begin{eqnarray}
{1 \over \cosh t} \left ( {\partial \over \partial r} + {\nu \over
\sin r} \right )  f
+ \left  ( M + i   { \partial \over \partial
t} +
  M  \right ) g  = 0 \, ,
\nonumber
\\
{1 \over \cosh t} \left ({\partial  \over \partial r}  - {\nu
\over \sin r} \right ) g
- \left  ( M + i  {\partial  \over
\partial t}  -    M \right ) f = 0     \, ;
\label{1.10a}
\end{eqnarray}

$\delta = -1$,
\begin{eqnarray}
{1 \over \cosh t} \left ( {\partial \over \partial r} + {\nu \over
\sin r} \right )  f
+ \left  ( M + i   { \partial \over \partial
t} - M  \right ) g  = 0 \, ,
 \nonumber
\\
{1 \over \cosh t} \left ({\partial  \over \partial r}  - {\nu
\over \sin r} \right ) g
 - \left  (M +  i  {\partial  \over
\partial t}  + M \right ) f = 0     \,;
\label{1.10b}
\end{eqnarray}

\noindent
 the term  $i\partial _{t} $  should be  neglected in comparing with  $2M$.
 Equations     (\ref{1.10a}),  (\ref{1.10b}) give respectively

$\delta = +1$,
\begin{eqnarray}
{1 \over \cosh t} \left ( {\partial \over \partial r} + {\nu \over
\sin r} \right )  f  + 2 M     g  = 0 \; ,
\nonumber
\\
{1 \over \cosh t} \left ({\partial  \over \partial r}  - {\nu
\over \sin r} \right ) g  -
 i  {\partial  \over \partial t}  f =\; 0     \; ;
\nonumber
\end{eqnarray}

$\delta = -1$,
\begin{eqnarray}
{1 \over \cosh t} \left ( {\partial \over \partial r} + {\nu \over
\sin r} \right )  f  +
 i   { \partial \over \partial t}  g  = 0 \; ,
 \nonumber
\\
{1 \over \cosh t} \left ({\partial  \over \partial r}  - {\nu
\over \sin r} \right ) g  - 2M  f =\; 0     \,.
\nonumber
\end{eqnarray}

In each case,  we derive an equation  for  the big component (the concomitant equation permits to
express the small component via   the big one)

\vspace{2mm}

$\delta = +1$,
\begin{eqnarray}
 i  {\partial  \over \partial t}  f = - {1 \over 2M}
 {1 \over \cosh^{2}  t}
\left ({\partial  \over \partial r}  - {\nu \over \sin r} \right )
 \left ( {\partial \over \partial r} + {\nu \over \sin r} \right ) f
     \; ,
\label{1.12a}
\\
g = -{1 \over 2M } {1 \over \cosh t}
\left ( {\partial \over
\partial r} + {\nu \over \sin r} \right )  f  \; ;
\nonumber
\end{eqnarray}

$\delta = -1$,
\begin{eqnarray}
i   { \partial \over \partial t}  g  =
 - {1 \over 2M} {1 \over \cosh t}
\left ( {\partial \over \partial r} + {\nu \over \sin r} \right )
 \left ({\partial  \over \partial r}  - {\nu \over \sin r} \right ) g
  \; ,
 \label{1.12b}
\\
f = {1 \over 2M} {1 \over \cosh^{2} t} \left ({\partial  \over
\partial r}  - {\nu \over \sin r} \right ) g
     \; .
\nonumber
\end{eqnarray}

Recall that the Pauli wave function is constructed from relativistic one in accordance with
\begin{eqnarray}
\varphi _{jm\delta } (x)\; =
 \left | \begin{array}{r}
     f_{1} D_{-1/2} \\
     f_{2} D_{+1/2} \\
\delta  f_{2} D_{-1/2}   \\
\delta  f_{1} D_{+1/2}
\end{array} \right |   \quad \Longrightarrow
\quad
\Psi^{Pauli}_{jm\delta } (x) =
 \left | \begin{array}{r}
     f_{1} D_{-1/2} + \delta  f_{2} D_{-1/2}\\
      f_{2} D_{+1/2} + \delta f_{1} \; D_{+1/2}
     \end{array} \right |.
 \label{1.13}
\end{eqnarray}

\noindent For different parities,  eq. (\ref{1.13}) gives respectively

\vspace{3mm}

$\delta = +1$,
\begin{eqnarray}
\Psi^{Pauli}_{jm, +1 }(t,r, \theta, \phi) 
=
 \left | \begin{array}{r}
     f_{1} D_{-1/2} +  f_{2} D_{-1/2}\\
      f_{2} D_{+1/2}  f_{1} D_{+1/2}
     \end{array} \right | =
     \left | \begin{array}{r}
     f (t,r) D_{-1/2}\\
     f  (t,r) D_{+1/2}
     \end{array} \right | ;
    \nonumber
\end{eqnarray}

$\delta = -1$,
\begin{eqnarray}
\Psi^{Pauli}_{jm, -1 } (t,r, \theta, \phi) 
=
 \left | \begin{array}{r}
     f_{1} D_{-1/2} - f_{2} D_{-1/2}\\
      f_{2} D_{+1/2} - f_{1} D_{+1/2}
     \end{array} \right |=
     \left | \begin{array}{r}
     -i g (t,r) D_{-1/2}\\
     -i g (t,r) D_{+1/2}
     \end{array} \right |.
 \nonumber
\end{eqnarray}

Note that according to  (\ref{1.12a}) and (\ref{1.12b}), non-relativistic functions  $f(t,r)$ and  $g(t,r)$
obey similar but slightly different equations:

\vspace{2mm}

$\delta = +1$,
\begin{eqnarray}
 i  {\partial f \over \partial t}= - {1 \over 2M}
 {1 \over \cosh^{2} t}
\left ({\partial  ^{2} \over \partial r^{2} }  - {\nu ^{2} + \nu \cos r \over \sin^{2} r} \right )
  f \,;
\label{1.15a}
\end{eqnarray}

$\delta = -1$,
\begin{eqnarray}
 i   { \partial g \over \partial t}=
 - {1 \over 2M} {1 \over \cosh^{2} t}
 \left ({\partial  ^{2} \over \partial r^{2} }  - {\nu ^{2} - \nu \cos r \over \sin^{2} r} \right )
  g \,.
\label{1.15b}
\end{eqnarray}

\noindent
In
(\ref{1.15a}), (\ref{1.15b}), the variables  are separated by the substitutions
\begin{eqnarray}
f(t,r) = f(t) f(r) \, , \quad g (t,r) = g(t) g(r) \, ;
 \nonumber
\end{eqnarray}

\noindent in this way we get

\vspace{2mm} $\delta = +1$,
\begin{eqnarray}
i   \cosh^{2} t {1 \over f(t) }{ d \over d t}  f (t)=
 - {1 \over 2M} {1 \over f(r)}
 \left ({d  ^{2} \over d r^{2} }  - {\nu ^{2} - \nu \cos r \over \sin^{2} r} \right ) f (r)= E
     \; ;
 \nonumber
\end{eqnarray}

$\delta = -1$,
\begin{eqnarray}
i   \cosh^{2} t {1 \over g(t) }{ d \over d  t}  g (t)=
 - {1 \over 2M} {1 \over g(r)}
 \left ({ d ^{2} \over d r^{2} }  - {\nu ^{2} - \nu \cos r \over \sin^{2} r} \right ) g (r)= E
    \,.
 \nonumber
\end{eqnarray}

The final separated equations are

\vspace{2mm}

$\delta = +1$,
\begin{eqnarray}
i   \cosh^{2} t {1 \over f(t) }{ d \over d t}  f (t) = E \quad
\Longrightarrow
\quad
 f(t)=e^{-iE\tanh t}\,,
\nonumber\\
 - {1 \over 2M} {1 \over f(r)}
 \left ({d  ^{2} \over d r^{2} }  - {\nu ^{2} + \nu \cos r \over \sin^{2} r} \right ) f (r)= E
     \; ;
 \nonumber
\end{eqnarray}

$\delta = -1$,
\begin{eqnarray}
i   \cosh^{2} t {1 \over g(t) }{ d \over d  t}  g (t) =E \quad
\Longrightarrow
\quad
 g(t)=e^{-iE\tanh t}\,,
\nonumber\\
 - {1 \over 2M} {1 \over g(r)}
 \left ({ d ^{2} \over d r^{2} }  - {\nu ^{2} - \nu \cos r \over \sin^{2} r} \right ) g (r)= E
    \,.
 \nonumber
\end{eqnarray}

Note the  symmetry between equations for $f(r)$ and $g(r)$:\  $\nu \Longrightarrow - \nu$;
therefore it suffices to examine in detail only one   case and  then to employ  the formal change.
For definiteness, we will consider the variant
 $\delta =+1$:
\begin{eqnarray}
 \left ({d  ^{2} \over d r^{2} }  - {\nu ^{2} + \nu \cos r \over \sin^{2} r}+2ME \right ) f (r) =0\,.
\label{1.18a}
\end{eqnarray}

\noindent Because the 3-subspace in de Sitter space-time  coincides with the compact spherical Riemann model,
 the motion if the variable $r$ must be quantized;  besides, we must assume that
  $2ME
> 0 $.
In eq.   (\ref{1.18a}), it is convenient to  use a new variable, $z = \cos r ,\;
z \in (-1, +1)$:
\begin{eqnarray}
(1-z^{2}) {d^{2} f \over dz^{2}} -z {df \over dz }
- \left (\nu{z
\over 1-z^{2}}+{\nu^{2} \over 1-z^{2}} -2ME \right )\, f = 0\,,
 \nonumber
\end{eqnarray}

\noindent and then to introduce  another variable
\begin{eqnarray}
y = {1-z \over 2} = {1 - \cos r \over 2} \;,
\nonumber\\
y(1-y)\,{ d^{2} f\over dy^{2}} + \left({1 \over 2} -y\right) \,{d
f\over dy}
- \left [{ \nu(\nu+1) \over 4 y} + {\nu (\nu-1) \over 4
(1-y)}-2ME   \right ] f=0 \; .
 \nonumber
\end{eqnarray}
With the substitution $f = y^{A} (1-y)^{B} F$, we get
\begin{eqnarray}
y(1-y){ d^{2} F\over dy^{2}} + \left[2A+{1 \over 2}
-(2A+2B+1)y\right] {d F\over dy}
\nonumber\\
- \left [(A+B)^{2}-{ 2A\,(2A-1)-\nu(\nu+1) \over 4 y} -
{2B\,(2B-1)-\nu (\nu-1) \over 4 (1-y)}-2ME   \right ] F=0 \; .
 \nonumber
\end{eqnarray}

\noindent At  restrictions
\begin{eqnarray}
2A = \nu +1, \;- \nu \, ;\quad  2B = -\nu + 1,\; + \nu \, ,
 \nonumber
\end{eqnarray}

\noindent the above equation simplifies
\begin{eqnarray}
y(1-y){ d^{2} F\over dy^{2}} + \left[2A+{1 \over 2}
-(2A+2B+1)y\right] {d F\over dy}
 - \left [(A+B)^{2}-2ME
\right ] F=0 \,,
 \nonumber
\end{eqnarray}

\noindent and coincides with the equation of hypergeometric type
\begin{eqnarray}
  y(1-y) \ F'' + [ (c - (a+b +1)y ]\ F' -ab \ F = 0 \, ,
 \nonumber
\end{eqnarray}
where
\begin{eqnarray}
c = 2A+{1 \over 2} \,,
\quad
 a+ b = 2A + 2B\, ,  \quad ab =
(A+B)^{2} -2ME \,;
 \nonumber
\end{eqnarray}
from whence it follows
\begin{eqnarray}
a =A+B - \sqrt{2ME} \, , \quad  b =A+B + \sqrt{2ME} \,.
 \nonumber
\end{eqnarray}

From physical considerations, we should construct solutions in polynomials
and they should be associated with discrete spectrum of energy
$
2ME > 0\; .
$
 Appropriate solutions are possible at positive  $A$ and
$B$:
\begin{eqnarray}
2A  = + \nu +1 = j+3/2 \,, \quad 2B  = + \nu  = j+1/2\,,
 \nonumber
\end{eqnarray}
polynomial condition provides us with the quantization rule
\begin{eqnarray}
a =  - n , \quad  n =0,1,2, ... \quad 2ME  = (j+ 1 + n)^{2}  \,
;
 \nonumber
\end{eqnarray}

\noindent corresponding radial functions are
\begin{eqnarray}
f (y) =  y ^{(\nu+1) /2} (1-y) ^{\nu /2}  F (a,b,c, y) \, ,
\nonumber\\
b =   2(j+1)  + n  \,, \quad c = j +2 \quad (\nu = j +1/2) \, .
 \nonumber
\end{eqnarray}

With the use of the mentioned  symmetry between  $f(r)$ and $ g(r)$,
one gets description of states with other parity
\begin{eqnarray}
g (y) =  y^{A'}(1-y) ^{B'} G (y) \,,
\nonumber\\
 2A' = - \nu +1,  \nu \,
, \quad   2B' = \nu +1,  - \nu \,  ;
 \nonumber
\end{eqnarray}
at positive  $A'$ and $ B'$:
\begin{eqnarray}
2A'  =  \nu  = j+1/2 \,, \quad  c' = j +1 =c-1\, ,
\nonumber\\
2B'  = + \nu +1  = j+3/2\,, \quad A'+B' = j +1 \, ,
\nonumber\\
a' =  - n=a , \quad    b' = 2(j+1)  + n  = b
\,,
\nonumber\\
g (y) =  y ^{\nu /2} (1-y)^{(\nu +1) /2}  F (a,b,c-1, y) \, .
 \nonumber
\end{eqnarray}

Thus, solutions of the Pauli equation in non-static de Sitter metrics
have been constructed

\vspace{2mm}

$\delta= +1$,
\begin{eqnarray}
\Psi^{Pauli}_{Ejm,+1} (t,r,\theta,\phi)
= e^{-iE\tanh t} f(r)
 \left | \begin{array}{r}
       D^{j}_{-m,-1/2}(\phi,\theta,0) \\
     D^{j}_{-m, +1/2}(\phi,\theta,0)
     \end{array} \right |,
\nonumber\\
f (r) =  \left (\sin {r \over 2}\right ) ^{j+3/2  } \left ( \cos{r
\over 2} \right )  ^{j+1/2 }
 F (-n,n +2j +2 ,j+2 ,
\sin^{2}{r\over 2} ) \; ;
 \nonumber
\end{eqnarray}

$\delta= -1$,
\begin{eqnarray}
\Psi^{Pauli}_{Ejm,-1} (t,r,\theta,\phi)
= e^{-iE\tanh t} g(r)
 \left | \begin{array}{r}
       D^{j}_{-m,-1/2} (\phi,\theta,0) \\
     D^{j}_{-m,+1/2}(\phi,\theta,0)
     \end{array} \right |,
\nonumber\\
g (r) =  \left (\sin {r \over 2}\right ) ^{j+1/2  } \left ( \cos{r
\over 2} \right )  ^{j+3/2 }
 F (-n,n +2j +2 ,j+1 , \sin^{2}
{r\over 2}) \; .
 \nonumber
\end{eqnarray}

Spectral parameter   $E$ (it does not represent the energy of stationary states) is quantized
according to the rule (the same for state with different parities)
\begin{eqnarray}
2ME  = (j+ 1 + n)^{2}\; .
 \nonumber
\end{eqnarray}

\section{Pauli equation in the  oscillating  anti de Sitter Universe}

In  the non-static anti de Sitter metric
\begin{eqnarray}
x^{\alpha } = ( t , r , \theta  , \phi  ),\quad\quad
  dS^{2} =  dt^{2}
  -  \cos^{2} t  [
dr^{2} + \sinh^{2} r ( d\theta ^{2} + \sin^{2} \theta  d\phi^{2}
)  ]
 \nonumber
\end{eqnarray}

\noindent and the corresponding diagonal tetrad
\begin{eqnarray}
e^{\alpha}_{(0)} = ( 1 , 0 , 0 , 0 ) \, ,
\quad
 e^{\alpha }_{(3)}
= \left ( 0 , { 1 \over   \cos t } , 0 , 0\right  ) \,
  ,
\nonumber\\
e^{\alpha }_{(2)} = \left ( 0 ,0 , 0 , {1 \over \cos t \, \sinh r \,
\sin \theta} \right )\, ,
\quad
 e^{\alpha }_{(1)} =\left  ( 0 , 0 ,{ 1 \over
\cos t\, \sinh r}, 0 \right )\, ,
 \nonumber
\end{eqnarray}

\noindent generally covariant Dirac equation  takes the form
\begin{eqnarray}
\left  [ i \gamma^{0}  \cos t  (\;  \partial_{t} \; - \;  {3 \over
2} \tan t  \; )+
 {1 \over \sinh r} \Sigma_{\theta \phi} + i \gamma^{3}  (   \partial_{r} \; + \;  {1 \over
\tanh r}  \; ) -  \; m  \cos t  \right ] \Psi  = 0\, .
 \nonumber
\end{eqnarray}

\noindent Separating a simper multiplier in
the wave function
\begin{eqnarray}
\Psi (x)  = {1 \over \sinh r} \,{1 \over \cos^{3/2} t } \; \varphi
(x) \,,
 \nonumber
\end{eqnarray}

\noindent we  obtain a more simple form
\begin{eqnarray}
\left [ i \gamma^{0}  \cos t  \;  {\partial \over \partial t} + i \gamma^{3}
{\partial \over \partial r} +
 {1 \over \sinh r} \Sigma_{\theta \phi}  -   m  \cos t  \right ]   \varphi  = 0\, .
 \nonumber
\end{eqnarray}

Using the technique of $D$-Wigner functions (\ref{1.4a})--(\ref{1.4c}), we obtain the radial equation
\begin{eqnarray}
i \cos t { \partial \over \partial t}    f_{3}
  -  i  {\partial \over \partial r}  f_{3}   - i {\nu \over \sinh r}
f_{4}  -  m \;\cos t  f_{1} =   0  \; ,
\nonumber\\
i \cos t { \partial \over \partial t}   f_{4} +  i  {\partial
\over \partial r} f_{4}   + i {\nu \over \sinh r} f_{3}  -  m
\;\cos t  f_{2} =   0    \; ,
\nonumber\\
i \cos t { \partial \over \partial t}   f_{1} +  i  {\partial
\over \partial r} f_{1}  + i {\nu \over \sinh r} f_{2}  -  m
\;\cos t f_{3} =   0  \; ,
\nonumber\\
i \cos t { \partial \over \partial t}    f_{2}
  -  i  {\partial \over \partial r} f_{2}
  - i {\nu \over \sinh r} f_{1}  -  m \;\cos t f_{4} =   0  \; .
 \nonumber
\end{eqnarray}

\noindent Using the operator of spatial parity
(\ref{1.6})--(\ref{1.8}),  we get more simple equations
\begin{eqnarray}
\left ( {\partial \over \partial r} + {\nu \over \sinh r} \right )
f
 + \left  ( i  \cos t \,{ \partial \over \partial t} +
 \delta  m  \cos t \right ) g  = 0  ,
\nonumber\\
\left ({\partial  \over \partial r}  - {\nu \over \sinh r} \right
) g  
- \left  ( i  \cos t\, {\partial  \over \partial t}  -
\delta  m  \cos t \right ) f = 0      .
\label{2.4}
\end{eqnarray}

Now, let us perform the nonrelativistic approximation in the system
 (\ref{2.4}).
First, we separate the rest energy. The cases  $\delta = \pm 1$ should be considered separately:

\vspace{2mm}

$\delta = +1$,
\begin{eqnarray}
{1\over \cos t}\left ( {\partial \over \partial r} + {\nu \over
\sinh r} \right ) f  + \left  ( M + i {\partial \over \partial t}
+
 M  \right ) g  = 0  ,
\nonumber\\
{1\over \cos t}\left ({\partial  \over \partial r}  - {\nu \over
\sinh r} \right ) g  - \left  ( M + i {\partial \over \partial t}
-
  M  \right ) f = 0      ;
 \nonumber
\end{eqnarray}

$\delta = -1$,
\begin{eqnarray}
{1\over \cos t}\left ( {\partial \over \partial r} + {\nu \over
\sinh r} \right ) f  + \left  ( M + i {\partial \over \partial t}
-
 M  \right ) g  = 0  ,
\nonumber\\
{1\over \cos t}\left ({\partial  \over \partial r}  - {\nu \over
\sinh r} \right ) g  - \left  ( M + i {\partial \over \partial t}
+
  M  \right ) f = 0     .
 \nonumber
\end{eqnarray}

\noindent These system  give respectively
 (the term  $i\partial _{t} $  should be  neglected in comparing with  $2M$)

\vspace{2mm}

$\delta = +1$,
\begin{eqnarray}
{1\over \cos t}\;\left ( {\partial \over \partial r} + {\nu \over
\sinh r} \right ) f  + 2M g  = 0 \; ,
 \nonumber\\
{1\over \cos t}\;\left ({\partial  \over \partial r}  - {\nu \over
\sinh r} \right ) g  - i {\partial \over \partial t}  f =\; 0
\; ;
 \nonumber
\end{eqnarray}

$\delta = -1$,
\begin{eqnarray}
{1\over \cos t}\;\left ( {\partial \over \partial r} + {\nu \over
\sinh r} \right ) f  + i {\partial \over \partial t} g  = 0 \; ,
\nonumber\\
{1\over \cos t}\;\left ({\partial  \over \partial r}  - {\nu \over
\sinh r} \right ) g  - 2M f =\; 0     \,.
 \nonumber
\end{eqnarray}

\noindent For both cases, we derive equations

\vspace{2mm}

$\delta = +1$,
\begin{eqnarray}
i {\partial \over \partial t}  f =-{1\over2M }\;{1\over \cos^{2}
t}
\left ({\partial  \over \partial r}  - {\nu \over \sinh r}
\right ) \left ( {\partial \over \partial r} + {\nu \over \sinh r}
\right ) f         \, ,
 \nonumber\\
 g=-{1\over2M }\;{1\over \cos t}\,\left ( {\partial \over \partial r} + {\nu \over \sinh r} \right )
f     \, ;
 \nonumber
\end{eqnarray}

$\delta = -1$,
\begin{eqnarray}
i {\partial \over \partial t} g =-{1\over 2M}\;{1\over \cos^{2}
t}
 \left ( {\partial \over \partial r} + {\nu \over \sinh r}
\right ) \left ({\partial  \over \partial r}  - {\nu \over \sinh
r} \right ) g    \, ,
\nonumber\\
f={1\over 2M}\;{1\over \cos t}\,\left ({\partial  \over \partial
r}  - {\nu \over \sinh r} \right ) g      \, .
 \nonumber
\end{eqnarray}

Nonrelativistic functions  $f(t,r)$ and  $g(t,r)$
obey similar  equations

\vspace{2mm}

$\delta = +1$,
\begin{eqnarray}
i {\partial \over \partial t}  f =-{1\over2M }\;{1\over \cos^{2}
t}
\left ({\partial  ^{2} \over \partial r^{2} }  - {\nu ^{2} + \nu \cosh r \over \sinh^{2} r} \right )
f           \, ;
\label{2.8a}
\end{eqnarray}

$\delta = -1$,
\begin{eqnarray}
i {\partial \over \partial t} g =-{1\over 2M}\;{1\over \cos^{2}
t}
\left ({\partial  ^{2} \over \partial r^{2} }  - {\nu ^{2} - \nu \cosh r \over \sinh^{2} r} \right ) g
   \, .
   \label{2.8b}
\end{eqnarray}

\noindent
In
(\ref{2.8a}), (\ref{2.8b}) the variables  are separated by the substitutions
\begin{eqnarray}
f(t,r) = f(t) f(r) \, , \quad g (t,r) = g(t) g(r) \, ,
 \nonumber
\end{eqnarray}

\noindent in this way we get

$\delta = +1$,
\begin{eqnarray}
i\,\cos^{2} t\,{1\over f(t)}\, {d \over d t}  f(t) =-{1\over2M
}\,{1\over f(r)}
\left ({d  ^{2} \over d r^{2} }  - {\nu ^{2} + \nu \cosh r \over \sinh^{2} r} \right )
f(r)=E           \, ;
\label{2.9a}
\end{eqnarray}

$\delta = -1$,
\begin{eqnarray}
i\,\cos^{2} t\,{1\over g(t)}\, {d \over d t} g(t) =-{1\over
2M}\,{1\over g(r)}
\left ({d  ^{2} \over d r^{2} }  - {\nu ^{2} - \nu \cosh r \over \sinh^{2} r} \right ) g(r)=E      \,.
\label{2.9b}
\end{eqnarray}

The final separated equations are

$\delta = +1$,
\begin{eqnarray}
i\,\cos^{2} t\,{1\over f(t)}\, {d \over d t}  f(t) =E\quad
\Longrightarrow
\quad
 f(t)=e^{-iE\tan t}\,,
\nonumber\\
-{1\over2M }{1\over f(r)}
 \left ({d  ^{2} \over d r^{2} }  - {\nu ^{2} + \nu \cosh r \over \sinh^{2} r} \right )
f(r)=E;
 \nonumber
\end{eqnarray}

$\delta = -1$,
\begin{eqnarray}
i\,\cos^{2} t\,{1\over g(t)}\, {d \over d t}  g(t) =E\quad
\Longrightarrow
\quad
 g(t)=e^{-iE\tan t}\,,
\nonumber\\
-{1\over 2M}{1\over g(r)}
 \left ({d  ^{2} \over d r^{2} }  - {\nu ^{2} - \nu \cosh r \over \sinh^{2} r} \right ) g(r)=E.
  \nonumber
\end{eqnarray}

Due to the  symmetry between equations for $f(r)$ and $g(r)$:   $\nu \Longrightarrow - \nu$;
it suffices to examine in detail only one   case and  then to employ  the formal change.
For definiteness, we will consider the variant
 $\delta =+1$:
\begin{eqnarray}
 \left ({d  ^{2} \over d r^{2} }  - {\nu ^{2} + \nu \cosh r \over \sinh^{2} r}+2ME \right ) f (r) =0\,.
\label{2.11a}
\end{eqnarray}

\noindent
In eq.   (\ref{2.11a}), it is convenient to  use the new variable $z = \cosh r ,\;
z \in (1, +\infty)$:
\begin{eqnarray}
(1-z^{2})\, {d^{2} f \over dz^{2}} -z\, {df \over dz }
- \left (\nu{z
\over 1-z^{2}}+{\nu^{2} \over 1-z^{2}} +2ME \right ) f = 0\,,
 \nonumber
\end{eqnarray}

\noindent and then to introduce  another variable
\begin{eqnarray}
y = {1-z \over 2} = {1 - \cosh r \over 2} \;,
\nonumber\\
y\,(1-y)\,{ d^{2}f \over dy^{2}} + \left({1 \over 2} -y\right) {d
f\over dy}
- \left [{ \nu\,(\nu+1) \over 4 y} + {\nu\,(\nu-1)
\over 4 (1-y)} +2ME  \right ] f=0 \, .
 \nonumber
\end{eqnarray}

\noindent
With the substitution $f = y^{A} (1-y)^{B} F$, we get
\begin{eqnarray}
y(1-y){ d^{2}F \over dy^{2}} + \left[{1 \over 2}+2A
-(1+2A+2B)y\right] {d F\over dy}
\nonumber\\
- \left [(A+B)^{2}-{2A\,(2A-1) -\nu\,(\nu+1) \over 4 y}
 -
{2B\,(2B-1)-\nu\,(\nu-1)  \over 4 (1-y)} +2ME  \right ] F=0 \, .
 \nonumber
\end{eqnarray}

\noindent At additional restriction
\begin{eqnarray}
 A
=-{1\over 2}\,\nu\,,\;{1\over 2}+{1\over 2}\,\nu \, ;
\quad
 B = {1\over 2}-{1\over
2}\,\nu\,,\;{1\over 2}\,\nu \,
 \nonumber
\end{eqnarray}

\noindent the above equation simplifies
\begin{eqnarray}
y(1-y){ d^{2}F \over dy^{2}} + \left[{1 \over 2}+2A
-(1+2A+2B)y\right] {d F\over dy}
- \left [(A+B)^{2}+2ME\right
] F=0 \, ,
 \nonumber
\end{eqnarray}
and coincides with the equation of hypergeometric type for $F(a,b,c,y)$ wuith
\begin{eqnarray}
c = 2A+{1 \over 2} \,,
\quad
a =A+B - i\sqrt{2ME} \, ,
\quad
  b =A+B + i\sqrt{2ME} \,.
 \nonumber
\end{eqnarray}

\noindent
 Corresponding solutions are
\begin{eqnarray}
a={1\over 2}+\nu- i \sqrt{2ME},
\quad
 b={1\over 2}+\nu+
i\sqrt{2ME},\quad
 c={3\over 2}+\nu = j +2 \,,
\nonumber
\end{eqnarray}
\begin{eqnarray}
a=1+j- i \sqrt{2ME}\,,\quad
 b=1+j+i \sqrt{2ME}\,,
\quad
 f (r) = C\, y ^{(\nu+1) /2} (1-y) ^{\nu /2}  F (a,b,c, y) \, .
 \nonumber
\end{eqnarray}

Allowing for the mentioned symmetry between  $f(r)$ and $ g(r)$,
one gets description of states with other parity:
\begin{eqnarray}
g =  C\,' \; y^{A'}(1-y) ^{B'} G\;,
\quad
 A' ={1\over 2}\,\nu\,,\;{1\over 2}-{1\over 2}\,\nu \, ,
\quad
 B' = {1\over 2}+{1\over 2}\,\nu\,,\;-{1\over 2}\,\nu \, ,
\nonumber\\
a'=1+j-i\sqrt{2ME}\,,
\quad
b'=1+j+i\sqrt{2ME}\,,\;\; c'=j+1=c-1\,,
\nonumber
\\
g (r) = C\,' \;  y ^{\nu /2} (1-y)^{(\nu +1) /2}  F (a,b,c-1, y)
\; .
 \nonumber
\end{eqnarray}

Thus, solutions of the Pauli equation in non-static anti de Sitter metrics
have been constructed
\begin{eqnarray}
\delta= +1, \qquad \Psi^{Pauli}_{Ejm,+1} (t,r,\theta,\phi)  = e^{-iE\tan t} f(r)
 \left | \begin{array}{r}
       D^{j}_{-m,-1/2}(\phi,\theta,0) \\
     D^{j}_{-m, +1/2}(\phi,\theta,0)
     \end{array} \right |,
\nonumber\\
f (r) = C\; \left ({1 - \cosh r \over 2}\right ) ^{(\nu+1)/2  }
\left ( {1 + \cosh r \over 2} \right )  ^{\nu/2 }  F \left(-n,\;n
+2\nu +1 ,\;\nu+{3\over 2} , \;{1 - \cosh r \over 2} \right) ;
 \nonumber
\end{eqnarray}
\begin{eqnarray}
\delta= -1, \qquad \Psi^{Pauli}_{Ejm,-1} (t,r,\theta,\phi)  = e^{-iE\tan t} g(r)
 \left | \begin{array}{r}
       D^{j}_{-m,-1/2} (\phi,\theta,0) \\
     D^{j}_{-m,+1/2}(\phi,\theta,0)
     \end{array} \right |,
\nonumber\\
g (r) = C'\; \left ({1 - \cosh r \over 2}\right ) ^{\nu/2  } \left
({1+ \cosh r \over 2} \right )  ^{(\nu+1)/2 }  F \left(-n,\;n
+2\nu +1 ,\;\nu+{1\over 2}, \;{1 - \cosh r \over 2}\right) .
 \nonumber
\end{eqnarray}

{\bf Acknowledgements}
\vspace{2mm}

 This  work was   supported   by the Fund for Basic Researches of Belarus,
 F 13K-079, within the cooperation framework between Belarus  and Ukraine,
 and by the Fund for Basic Researches of Belarus,
 F 14ARM-021, within the cooperation framework between Republic of Belarus  and  Republic of Armenia.

The authors are grateful to V.M. Red'kov for advice and help.

\end{document}